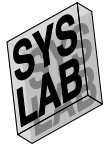

# Systems, Views and Models of UML[*]

Ruth Breu, Radu Grosu, Franz Huber,
Bernhard Rumpe, Wolfgang Schwerin

email: {breur,grosu,huberf,rumpe,schwerin}
@informatik.tu-muenchen.de

Technische Universität München
Arcisstr. 21
D-80290 München, Germany

**Abstract**

In this paper we show by using the example of UML, how a software engineering method can benefit from an integrative mathematical foundation. The mathematical foundation is given by a mathematical system model. This model provides the basis both for integrating the various description techniques of UML and for implementing methodical support. After describing the basic concepts of the system model, we give a short overview of the UML description techniques. Then we show how they fit into the system model framework and sketch an approach to structure the UML development process such that it provides methodological guidance for developers.

# 1  Introduction – Why Formalization?

"The nice thing about graphical description techniques is that everyone understands them, the bad thing, however, is that everyone understands them in a different way." This often heard quote captures a main property of modeling techniques using semi-formal, mostly graphical notations, beginning with the early structured modeling techniques and prevalent until today's object-oriented methods. The diagrammatic notations used here seem easily comprehensible for everyone dealing with software development. Experience shows, however, that, except from obvious properties, these notations do in fact bear a great number of possible divergent interpretations. To gain a deeper and more exact understanding of the notations used, the need for providing a stringent formal foundation for them has long been recognized and has lead to considerable advances in the recent past. Many of these efforts aim at providing a formal semantics for single notations. Considering the different aspects of a system, described by different notations, captured in a system development process, providing isolated semantics, possibly even using diffent models as a basis, does not seem to be adequate. The complete description of such a system is given only by the assembly of all different views. It is therefore desirable to have a common semantic basis for *all* notations used in a development method. This allows both to provide an exact interpretation for diagrams in a single notation and to inter-relate diagrams in different notations on a common basis.

---

[*] This paper originates from the SysLab project, which is supported by the DFG under the Leibnizprogramme and by Siemens-Nixdorf.

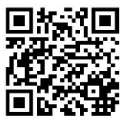

The SysLab method (Breu *et al.* (1997)b) builds upon such an integrated semantic framework, which we call the *System Model*. The foundations and the basic mathematical concepts of this system model will be introduced in Section 2. The SysLab method, just like the Unified Method, is a view-oriented, integrated, and object-oriented development method. Its description techniques are deliberately similar to those used in the UML. Thus it seems worthwhile to apply SysLab's system model to the UML in order to embed its description techniques into the system model's mathematical framework. What can be gained by this effort is, quite obviously, a deeper understanding of the single notations and, as outlined above, the possibility to more tightly intergrate the UML description techniques on a sound mathematical basis. The benefits hereof can be enormous, especially for tool developers, resp. tool vendors, and methodologists. Tool developers are enabled to provide a much larger set of precise consistency conditions than currently possible by using just the UML *Meta Model* which basically represents only the abstract syntax of UML's description techniques. Using the semantic properties of diagrams produced in different stages of a development project and their inter-relationships, transformation tools can be provided, which tranform, either automatically or with a human developer's assistance, documents from one notation into another. On this basis, methodological guidelines for software developers can be elaborated and eventually implemented in CASE tools.

The rest of the paper is organized as follows. In Section 2 we present the basic concepts of the SysLab system model and in Section 3 we give a brief overview of the UML description techniques. We then outline methodological aspects in Section 4, where we define general properties of a software development method and show how a development process using UML description techniques can be structured and supported by a tool. Finally, an outlook on further steps in formalization and tool support is given in Section 5.

# 2 Models of Systems

In general, a *system* is a technical or sociological structure consisting of a group of elements combined to form a whole and to work, function or move interdependently and harmoniously. A *system model* represents certain aspects of systems in a certain way, using certain concepts, e.g. OO-concepts, such as classification, encapsulation etc. . One way to formulate system models is to use mathematical techniques (e.g. sets, functions, relations, predicates). This leads to the notion of *mathematical* system models.

In the following we first motivate the use of a system model and then describe the one on which the SysLab method is based and which is also appropriate for UML.

## 2.1 Motivation of the System Model

In the introduction we have motivated, *why* a formalization of UML description techniques is useful. We argued that a precise semantics is important not only for the developer, but also for tool vendors, methodologists (people that create the method) and method experts (people that use the method and know it in detail).

We get the following requirements for a formalization:

1. A formalization must be complete, but as abstract and understandable as possible.

2. The formalization of a heterogeneous set of description techniques has to be integrated to allow the definition of relations between them.

This does not mean that every syntactical statement must have a formal meaning. Annotations or descriptions in prose are always necessary for documentation, although they do not have a formal translation. They may however eventually be translated into formal descriptions or even into code in the course of a software development process.

As soon as one uses description techniques with a fixed syntax and semantics, one no longer describes systems in general but models of systems. Using UML, for example, one models systems in terms of objects, classes etc. To manage the complexity of formalization, we introduce a layer between the syntactic description techniques and pure mathematics as shown in Figure **??**.

This intermediate layer is the mathematical system model. On this layer, various aspects of systems such as structure, behavior and interaction are formulated by mathematical techniques. We call the representation of a system, formulated solely in terms of the system model, a *model* of a system. Furtheron, the set of all models that can be described in terms of a certain system model is called the *universe* of this system model. Figure **??** illustrates the ideas that are explained below.

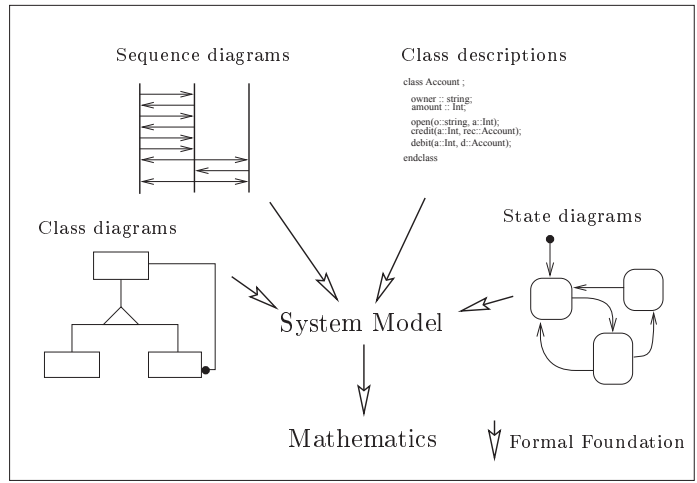

Figure 1: Layered formalization of description techniques

Description techniques offer syntactical elements that allow the specification of certain views, i.e., certain aspects, of systems. One way to define a semantics is to express the meaning of syntax in terms of a system model. Using a unique system model for a set of description techniques results in an integrative semantics for the different techniques. An integrative semantics allows reasoning about interrelations between different views expressed by different techniques. Notions like *refinement*, *consistency* and *completeness* can be precisely defined.

A *mathematical* system model provides terms with a formal semantics, e.g. functions, and can therefore serve as a basis for a *formal* semantics for a set of description techniques.

With the above definitions, one can define the semantics of a *document* (which is a kind of a module on the syntactical level) of a given description technique as the set of all those models, that have the properties that are expressed in the document.

Using a set of models and not a single one as the basis of the proposed semantics has several advantages. For example, refinement of documents corresponds to set inclusion. Furthermore, we get the meaning of different documents modeling different aspects of the system by intersection of their respective semantics. But the main reason is that, in contrast to fully executable programming languages, description techniques allow *underspecification* of system properties in many different ways. A proper semantics cannot be therefore captured by a single model. For the same reason, it is not possible to give an operational semantics in the sense that a document specifies a single abstract machine that "executes" it.

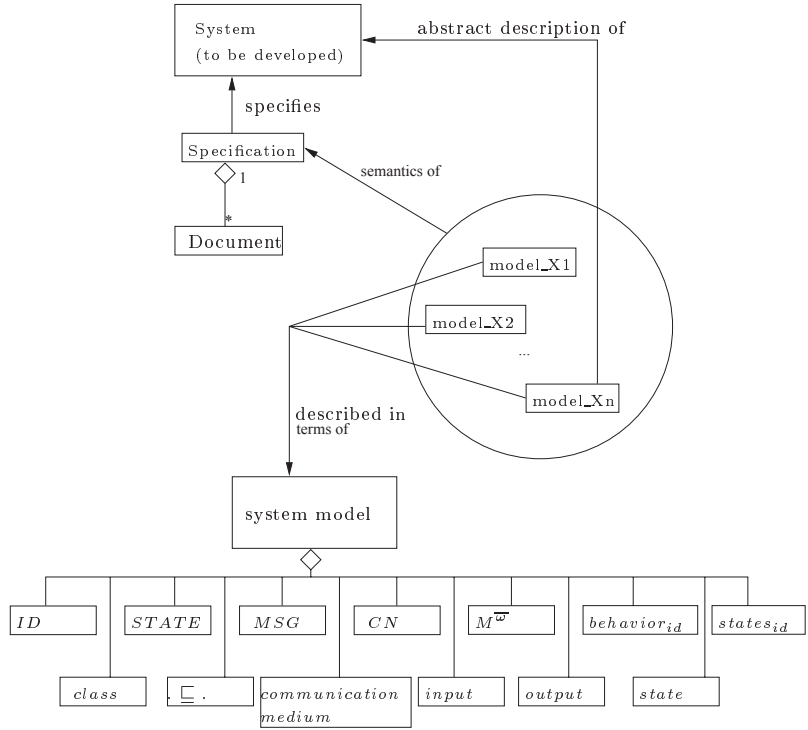

Figure 2: System Model, Models and Specification

## 2.2 Definition of the System Models

The system characterisation given below is a refinement of the SysLab system model as presented in Klein *et al.* (1996) and Grosu *et al.* (1996), and it is rather similar to the one used in Breu *et al.* (1997)b. Each document, for instance an object diagram, is regarded as a constraint on the system model's universe.

The system model introduced below is especially adapted for the formalization of UML. Thus, relevant aspects of UML like classes, objects, states, messages etc. are explicitly included. A precise formalization of our UML system model is currently under development.

Our system model is very general, covering various kinds of object-oriented systems such as conventional object-oriented software systems, systems including hardware components, embedded systems and real-time systems. For the formalization of the current set of UML notations, we will only need a specialized version of the system model, which is briefly presented below.

In the following, we describe the most important elements of our system model.

The structure of a system is, according to object-orientation, given by a set of objects, each with a unique identifier. Therefore, we regard the enumerable set *ID* of object identifiers as an element of each model.

In any system objects interact by means of *call message passing*. By this term we express that objects on the one hand communicate via a sequential call-return mechanism, but on the other hand have the possibility to send messages asynchronously, which means that the receiver may neither deny messages nor block the sender of a message. Both communication mechanisms are treated within one framework in detail in Paech & Rumpe (1997).

Asynchronous communication models provide the most abstract models for systems with message exchange, since deadlock problems as in synchronous systems do not occur. To model communication between objects we use the theory of timed communication histories as given in Broy *et al.* (1993). The notion of explicit time in the system model allows us to deal with real time, as proposed in UML.

We regard our objects as spatially or logically distributed and as interacting in parallel. As described in UML, sequential systems are just a special case, where always exactly one object is "active".

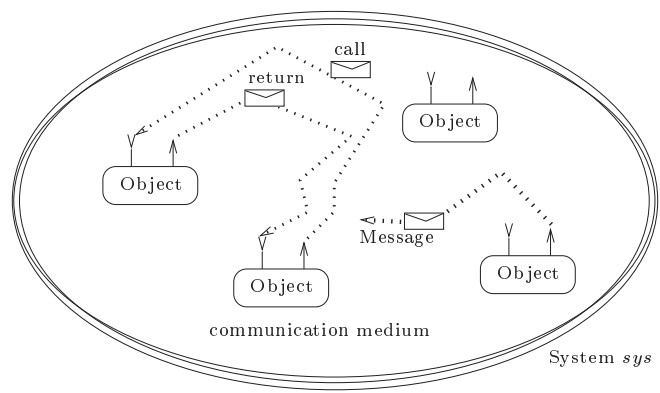

Figure 3: Objects in the UML system model

Interaction between objects occurs through the exchange of messages, as shown in Figure 3. Let *MSG* be an element of the system model, denoting the set of all possible messages in a system. An object with identifier $id \in ID$ accepts a unique set of messages. Its input interface is defined by

$msg_{id} \subseteq MSG$

The *behavior* of an object is the relationship between the sequences of messages it receives and the sequences of messages it emits as a reaction to incoming messages. We allow our objects to be nondeterministic, such that more than one reaction to an input sequence is possible.

According to Broy *et al.* (1993); Broy & Stølen (1994), the set of timed communication histories over $M$ is denoted by $M^{\overline{\omega}}$. A communication history is basically an infinite sequence containing (possibly only a finite number of) messages and time stamps inbetween, that mark time progress. Thus the messages occurring in a communication history are in linear order. A communication history models the observable sequence of incoming or outgoing messages of one object. The behavior of a nondeterministic object $id$ is then given by the mapping of its input stream to the set of possible ouput streams. Using relations, the behavior of an object $id$ is given by the relation between its input and output streams

$$behavior_{id} \subseteq msg_{id}^{\bar{\omega}} \times MSG^{\bar{\omega}}$$

Objects encapsulate data as well as processes. *Encapsulation of data* means that the state of an object is not directly visible to the environment, but can be accessed using explicit communication. *Encapsulation of process* means that the exchange of a message does not imply the exchange of control: each object can be regarded as a separate process. Note, that this view on object controls also works in conventional sequential programs. Objects get active when receiving a message and fall asleep after emitting a message. Given the set of possible states $STATE$ of objects in a system, the function *states* assigns a subset of possible states to every object $id$:

$$states_{id} \subseteq STATE$$

Furthermore, a state transition system is associated with each object, modeling the connection between the behavior and the internal state of an object. We use a special kind of automata (Grosu & Rumpe (1995)) for this purpose. An automaton of an object with identifier $id$ consists of a set of input messages $msg_{id}$, a set of output messages $MSG$, a set of states $states_{id}$, and a set of initial states $states_{id}^0 \subseteq states_{id}$. The nondeterministic transition relation $\delta_{id}$ defines the behavior of the automaton. From the state-box behavior, given for the automaton in terms of state transitions, the black-box behavior in terms of the *behavior*-relation can be derived (Grosu & Rumpe (1995)).

Messages are delivered by a *communication medium*, which is an abstraction of message passing as it is done in real systems by the runtime system of the programming language in combination with the operating system. The communication medium buffers messages as long as necessary. Each message contains the receiver's identifier, so that the communication medium contains of a set of message buffers, one for each object. The order of messages between two particular objects is always preserved by the communication medium. The contents of messages are not modified. Messages cannot be duplicated or lost. No new messages are generated by the communication medium. This is formalized in Grosu *et al.* (1996).

Each system allows a possible set of system runs. A system run is characterized by the messages exchanged between all the objects and the sequence of their states. We thus characterize a system run by the following three functions:

$$input : ID \to MSG^{\bar{\omega}}$$
$$output : ID \to MSG^{\bar{\omega}}$$
$$state : ID \to STATE^{\bar{\omega}}$$

associating with each object identifier the stream of messages the object receives in a run, the stream of messages the object emits in a run, and the stream of states the object has during the run. Of course, this trace like view on a system is strongly interconnected with the automaton based view given previously. The use of this trace like view is only possible, if objects are regarded as process capsules and therefore each computational step equals one step of an automaton (of the above characterized type) for exactly one object. Furthermore, as objects don't share common variables, computational steps of different objects cannot interfere and can therefore be serialized.

Objects are grouped into classes. We assume that each system owns a set $CN$ of class names. $CN$ may, for instance, be derived from UML class diagrams. In object-oriented systems, each object identifier denotes an object that belongs to exactly one class. This fact is represented by the function

$\quad class : ID \to CN$.

Classes are structured by an inheritance relation, which we denote by $. \sqsubseteq .$ (read: "subclass of"). The inheritance relation is transitive, antisymmetric and reflexive, as usual. With every class $c \in CN$ a signature $\Sigma_c$ is associated, containing all attributes and methods together with their argument and result types. The signature induces a set of input messages and a set of states for each object of the class. One impact of inheritance is that signatures are only extended: $c \sqsubseteq d \Rightarrow \Sigma_d \subseteq \Sigma_c$.

Another concept of object-orientation is the dynamic creation of objects. Deletion need not be modeled, as we assume that our objects are garbage collected in the usual way. However, we may define a special *finalize*()-method that may be used to clean up objects, as, for instance, in Java.

Initially, a finite subset of objects (usually one main-object) exists and is active. We regard all other objects to be created in the course of a system run and to be active after having received a first message. Thus, the creation of a new object essentially consists of a message transmission from the creator to the created object. To allow this, each object is equipped with a sufficiently large (usually infinite) set of object identifiers denoting the set of all object identifiers the object may create:

$\quad creatables : ID \to \mathcal{P}(ID)$

To prevent multiple creation, these sets of identifiers have to be pairwise disjoint, and objects that are initially active are not creatable at all.

## 3 Views

In the following section, we explore the meaning of a "view". Afterwards, we briefly explore which UML notation describes which aspects of a system.

### 3.1 Views and projections

A view of a system is a projection of the system on one of its relevant aspects. Such a projection emphasizes certain aspects and ignores others. Therefore it is useful to have different views of a system. This also allows each stage of the development to concentrate on relevant aspects and to delay others, that are at the moment less important. In general different kinds of projections

can be found, that are rather orthogonal and therefore constitute different dimensions:

- Projection on development phases: The same "thing" may have different appearances during analysis, design and implementation phases, and therefore described by different notations, or even not appear at all.

- Projection on structural, behavioral, interface and data aspects.

In general a projection can be any combination of the above mentioned kinds.

A document describes such a projection. As each document is of a certain kind, it projects certain aspects of the system, that can be described within the document. UML therefore uses multiple notations, that focus on different aspects, therefore exhibiting different "views" of the system.

Basically there are four main views:

- The *structural view* focuses on the structure of a system. It describes layout between objects and classes, their associations and their possible communication channels.

- The *behavioral view* focuses on the behavioral aspect of the system components. It describes how they interact, and characterizes the response to external system operations.

- The *data view* focuses on the data aspects of the system. It describes the state of the system units (objects), as well as their relationships.

- The *interface view* focuses on the encapsulation of system parts, and the possible usage from outside, e.g., through characterizing signatures.

Although these four views focus on different aspects, there are close relationships between them. Therefore it can be expected, that the notations, describing these views are also related, and that there are context conditions between them. This is currently one of the major problems of UML.

## 3.2 Notations of UML

UML currently has as many as eleven different and partly overlapping notations, which constitute different views of UML designs. In the following, we will discuss briefly, which notation can be used to describe certain aspects, without introducing the notations (this can be found in one of the various UML books, e.g. Burkhardt (1997); Booch *et al.* (1996)).

**Class diagrams** are the central notation for structural aspects. They define classes, their associations and how they are aggregated. However it is also possible to add data information, as each class can have an attribute section. Furthermore the connection to behavioral aspects is given by a method signature section, that can also be attached to classes.

Besides defining structural aspects, a class diagram may be used as source of data information, when transforming it into a database schema definition.

**Object diagrams** describe the actual layout of a part of the system within a certain situation. They clearly focus on structure, partly on contained data, and have to be compatible with the class diagrams. They can serve as a basis for behavioral descriptions.

**Packages** group classes (resp. their implementations) together. Their main focus is the definition of interfaces. They also define structure, but on a different level than class or object diagrams do: A package structures interfaces and implementations and is important during development. Relationships between packages are usually rather independent from the relationships between their included classes. Also due to the dynamics of object-orientation, this is again, to some extent, independent of the structure of the instantiated objects. Therefore we have different levels of structural views to be served.

**Use case diagrams** show the relationships among actors and use cases within a system. Although the concept of use cases seems to be very helpful, the actual use of this notation is to some extent unclear and will heavily depend on the method, which is still in development. However, use cases deal with interface and behavioral aspects at the border of the system. They exhibit possible actions to be taken, and who is allowed to undertake these actions. This means that there is also some structural aspect within the current UML use case notation.

**Sequence diagrams** describe a pattern of interaction among objects. Interactions between participating objects are arranged in timed sequences similar to Message Sequence Charts (MSC). Sequence diagrams therefore clearly define behavioral aspects but are based on structural and interface views.

**Object lifelines** are somewhat similar to sequence diagrams, but focus more on the control structure of an object and its related thread. They deal with the lifetime of a single object. They are used for describing behavioral aspects of a single object and are therefore more implementation oriented than ordinary sequence diagrams.

**Collaboration diagrams** are based on object diagrams, exhibiting a certain (numbered) flow of messages between objects in order to describe interaction between the participants. Collaboration diagrams and sequence diagrams are very similar in content, and it seems it is to some extent a matter of taste, which notation is to be preferred. Collaboration diagrams therefore also focus on behavior pattern.

Design patterns as introduced in UML comprehend a compact notation for collaboration patterns and are incorporated into class and object diagrams.

**State diagrams** are the central notation for describing behavior of a single object. This behavior description is based on the state space the object has, and it is also related to the message interface. State diagrams are therefore the central notation to relate data aspects and behavioral aspects of objects. It was therefore natural to enhance state diagrams with many features for different purposes. Examples are hierarchical structuring of the state space, state activities, entry and exit actions, state dependent attributes, or the history mechanism. Also hierarchical structuring of the event set and complex transitions have been defined. However, these newly added concepts interact with each other in ways, that have not yet been fully explored, and it seems advisable not use them too much.

**Activity diagrams** are defined as "a special case of a state diagrams" (Booch *et al.* (1996)), where the occurring states are named as "action states". However, there are serious doubts about that. It could be wise to regard activity diagrams as a form of data flow diagrams with additional components for control flow. Therefore activity diagrams could be useful for describing internal processing of operations (or use cases). Thus they clearly focus on behavioral aspects of individual components, but more on its functional decomposition into different actions.

Swimlanes in activity diagrams, furthermore allow the decomposition and regrouping of a series of actions for implementation by different objects.

Other notations exist, that are more implementation oriented, and thus mainly deal with the physical structure of a system. These notations only partly deal with some of the above mentioned views:

**Component diagrams** show the dependencies among software components.

**Deployment diagrams** show the configuration of runtime processing elements, processes and the objects that live on them.

# 4 Methodological Aspects

At the current stage, UML is not more than a syntactic framework for system specification. What is missing to make UML to a full-fledged software engineering method is what the word "method" is characterizing: a set of rules that guide the designer to obtain a runnable and correct system implementation.

In this section we will discuss in more detail what aspects a method should cover and identify three kinds of methodological rules (subsection 4.1). Subsection 4.2 sketches our view of the design process and its interconnection with the methodological rules. We give an idea of document graphs documenting design decisions and the dependencies between the specifications developed during design.

## 4.1 What a method is

It is not easy to define what a method is and we do not want to give an exact definition here. It is however certain that a method consists, apart from the notations themselves, of a set of rules and steps guiding through the *design process*. The design process is generally defined to be the sequence of specifications developed during the lifetime of the software system (including documents specifying extensions and modifications of the system).

The design process for large systems is very complex and thus the methodological rules deal with many different aspects and activities during design. In order to come closer to the kernel aspects of a method, we distinguish rules at three levels of abstraction,

- process models
- procedural guidelines and
- technical steps. [1]

**Process Models.** Rules on a high level of abstraction structure the design process as a whole. Typically they define *design phases* and the kind of documents that have to be produced in these phases. We call a set of such rules a *process model*.

Examples of process models are the classic waterfall model or the spiral model (cf. Boehm (1994)). A characteristic of these models and of process models in general is that they are independent of notations and even independent of the underlying system view (function oriented or object oriented).

Nevertheless it can be observed that process models in an object oriented setting follow a different design philosophy than the classic models and support a more activity oriented design process than a phase oriented design process. Such activities during the development are analysis, implementation and testing, but also prototyping and reuse. In this way, the design process becomes more flexible and adaptable to particular needs.

**Procedural Guidelines.** Process models are mainly concerned with the question of *when* particular specifications have to be produced. In addition,

---

[1] The latter two terms have been taken from Hussmann (1994), however with a slightly different meaning

rules at a lower level of abstraction support the designer *how* to produce a specification. These kinds of rules are in most cases of a heuristic, informal nature. They are formulated in the terminology of the underlying system view but independently of particular notations (e.g. types of diagrams). In the sequel we will call these rules *procedural guidelines*.

As an example, many object oriented methods give procedural guidelines to support the designer in identifying objects, methods and attributes in the application domain. Moreover the paradigm of design with use cases also comes along with a set of procedural guidelines.

**Technical Steps.** Rules at the lowest level of abstraction are both dependent of the system view and of the notations used. We call them *technical steps*.

Technical steps deal with the transformation of documents during the design process and with the interrelation of different notations. As an example, technical steps give rules for the refinement of diagrams or rules that ensure the consistency of a specification. Technical steps thus support the designer's understanding of the use of the description techniques and the production of a sound and complete system specification.

In contrast to their importance, technical steps are the most neglected kind of rules in today's methods, since they require a deep understanding of both the system view and the description techniques. While process models and procedural guidelines are less amenable to a formal treatment due to their heuristic nature, a formal foundation of the system view and the notations like our model is an eminent basis for studying technical steps in a systematic way.

In subsection 4.2 we sketch our view of the design process and clarify the semantic treatment of technical steps within the semantic model. Before, table 1 summarizes the concepts discussed in this section.

## 4.2 A View of the Design Process

A system specification in UML consists of a set of documents like sequence diagrams, class diagrams or free text. In order to support the effective management of a large number of documents, we extend this simple view and consider a system specification to be a *document graph* containing the following kind of information.

- The nodes of the document graph are UML documents.

- Each document in the graph has a *state* documenting its stage of development.

- A set of *relationships* between nodes describe dependencies between documents.

**Document state.** The document state is intimately connected with the notion of a document lifecycle. Apart from information, by whom and when some document has been created or updated, the document state reports whether the document has been *validated* (in case of informal and formal documents), *verified* (in case of formal documents) or *tested* (in case of programs).

Further attributes determining the state of a document are the *redundancy* of a document or its *consistency*. A document is redundant if all the properties of the system it describes can be derived by other documents in the graph.

**Document relationships.** Documents in UML have been syntactically decoupled as far as possible in order to support a flexible use during design. Thus, it is the meta level that has to document the structure of specifications. There are two main kinds of relationships in the document graph.

The first one is a kind of import or clientship relation. We call it the **refers to** relation. The **refers to** relation is directed and relates a document with the set of documents that have to be known in order to understand the document. For instance, a sequence diagram always has to refer to some class diagram in order to understand the labels of the object lifelines. As a further example, a state diagram may refer to a class diagram or to a method specification.

The second kind of relationship between documents is the **transform** relation. Also the **transform** relation is directed and relates specifications that are involved in some transformational step during design. A document $A$ transforming a document $B$ is the result of some technical step and describes some aspect of the system in more detail than document $B$. More generally, a technical step and the **transform** relation, respectively, may be based on more than one document and may produce several documents.

In our semantic framework, the **transform** relation has its counterpart in a semantic refinement relation on the semantic domain of documents. The refinement relation guarantees that the system properties specified in earlier stages are preserved during the development. Hence the technical steps formulate conditions at the syntactic level of UML that ensure that the refinement relation holds at the semantic level. The rules of the technical steps have to be formulated for every kind of document and have to be proved correct with respect to the semantic refinement relation.

In the current stage of development we have developed technical steps for the development of class diagrams and state diagrams (Rumpe (1996)). Rules concerning the interrelation between sequence diagrams and state diagrams are currently under investigation (Breu *et al.* (1997)a). Table 1 sketches the main idea of these technical steps. Figure 4 depicts a sample document graph. It has to be stressed again that the depicted diagram is *not* an UML document itself but a concept at the meta level, e.g. produced and administrated by a tool. Moreover, document graphs are rather used as a paradigm for managing the design process than as a kind of document suitable for graphical representation.

Integrating the transform relation, a document graph does not only represent a system specification at a single stage of development but represents the whole design process. Compared to the traditional view of the design process as a sequence of system specifications the document graph view is advantageous for several reasons.

| |
|---|
| **Source of transformation**: Class diagram(s) |
| **Produced documents**: Class diagram |
| **Description** |
| Allowed activities for the design of class diagrams are |
|     the introduction of new classes, attributes and methods, |
|     the introduction of new associations and inheritance relations, |
|     the strengthening of invariants and |
|     the integration of several class diagrams. |
| **Source of transformation**: State diagram |
| **Produced documents**: State diagram |
| **Description** |
| Allowed activities for the design of state diagrams are |
|     the introduction of new states, |
|     the refinement of states, |
|     the introduction of new transitions under certain conditions, |
|     the deletion of transitions. |
| The exact rules can be found in Rumpe (1996). |
| **Source of transformation**: Sequence diagram(s) |
| **Produced documents**: State diagram(s) |
| **Description** |
| Sequence diagrams describe exemplary event traces that may be synthesized to complete descriptions in state diagrams. Sequence diagrams roughly correspond to paths in the synthesized state diagram. |

Table 1: Technical Steps for UML documents

First, the notion of document graphs supports a design which does not have to be strictly phase oriented and homogeneous but enables the prototyping of subsystems and the reuse of documents. In this respect, the notion of document graphs is particularly suitable in an object oriented design environment.

Second, a system specification in any case consists of documents at different levels of abstraction (e.g. for documentation or communication purposes). Thus, the representation of the whole design process in the document graph is only a matter of consequence. It is clear that an explicit versioning concept for documents is not needed in our framework since the sequence of design steps is represented by a sequence of transform relations in the document graph.

# 5 Outlook and Further Work

In this paper, we have outlined directions for a stronger integration of the description techniques provided by the Unified Modeling Language. The underlying basis used for the integration is a mathematical model developed in the SysLab method, the *system model*.

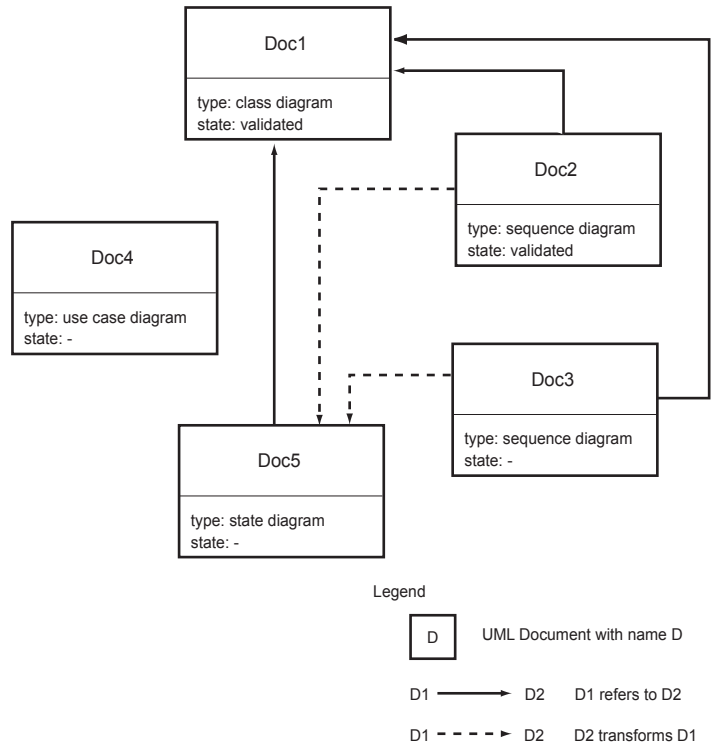

Figure 4: A Sample Document Graph

This setting provides a rich field for future research activities. As a start, precise mappings of the UML description techniques onto the mathematical system model have to be defined. Based on them, notions of consistency between development documents of the same or different description techniques can be defined. Consequently, a further step of research would be to use the interrelationships between the UML description techniques defined on the basis of the system model to soundly integrate them methodically, particularly on the level of the technical steps introduced in Section 4.1. The last step, building upon such an integration framework, would be to develop appropriate tools that support these techniques and the methodology.